\documentclass[10pt,endnotes]{usetex-v1-tight}
\usepackage{epsfig,endnotes}
\usepackage{url,subfiles,xcolor,pgfplots,amssymb,amsmath,subcaption,balance}

\begin{document}

\newcommand{\K}[1]{\noindent\textcolor{blue}{\bf $\blacksquare$ K: #1}}
\newcommand{\V}[1]{\noindent\textcolor{orange}{\bf $\blacksquare$ V: #1}}
\newcommand{\D}[1]{\noindent\textcolor{red}{\bf $\blacksquare$ D:: #1}}
\newcommand{\R}[1]{\noindent\textcolor{olive}{\bf $\blacksquare$ R: #1}}

\newcommand{\ITEM}[1]{\noindent\textcolor{black}{\bf $\blacksquare$~#1}}



\title{\Large \bf Towards Marrying Files to Objects \vspace{5pt}}
\author{
        {\rm Kunal Lillaney$^{1}$,
                Vasily Tarasov$^{2}$,
                David Pease$^{2}$,
                Randal Burns$^{1}$
	}\\
        {\rm {$^1$}{John Hopkins University}, {$^{2}$}{IBM Research---Almaden}}
}


\maketitle

\pagestyle{plain}

\subsection*{Abstract}

To deal with the constant growth of unstructured data,
vendors have deployed scalable, 
resilient, and cost effective object-based storage systems built on RESTful web services.
However, many applications rely on richer 
file-system APIs and semantics, and cannot benefit from object stores.
This leads to storage sprawl, as object stores are 
deployed alongside file systems and data is accessed and managed across both 
systems in an ad-hoc fashion.
We believe there is a critical need for a transparent merger of objects
and files, consolidating data into a single platform.  
Such a merger would extend the capabilities of both object and file stores
while preserving existing semantics and interfaces.
In this position paper, we examine the viability of unifying object stores
and file systems, and the various design 
tradeoffs that exist. 
Then, using our own implementation of an object-based,
POSIX-complete file system, we experimentally demonstrate
several critical design considerations.

\section{Introduction}
\label{section:introduction}

Object storage has gained significant traction in recent years owing to an
explosion in the amount of unstructured data and the popularity of applications
using RESTful interface to access such data.
Some estimates predict that by 2019 more than 30\% of the storage capacity in 
data centers will be provided by object storage~\cite{Chandrasekaran:2017}, 
and that by 2020 the world will produce 44 zetabytes of
unstructured data each year~\cite{Turner:2014}.
Object stores can offer cost of ownership and scalability
improvements over traditional file systems.
Users and enterprises increasingly look to object storage
as an economical solution
for storing unstructured data~\cite{Cockroft:2011}~\cite{Deelman:2008}.
Object stores are characterized by Web-style access, flat namespaces,
immutable data, relaxed consistency,
and rich user-defined metadata.  
%
Most object stores support simple data access operations:
\texttt{GET}, \texttt{PUT}, and
\texttt{DELETE}.
%

Existing applications are often unable to benefit from object storage because 
they rely on traditional file system interfaces and features.
%
%
Many applications, for example 
genomic workflows~\cite{O:2013}~\cite{Langmead:2010}, 
are dependent on a namespace hierarchy and file pointers 
which are not supported by object storage.
Objects do not support updating in place,
and need to be updated and rewritten as a unit;
existing big data workflows may need to be rearchitected to use object 
storage in 
order to avoid the performance penalty introduced by incremental 
updates~\cite{Vahi:2013}.
Given the flat name space in object stores, some space organization 
operations such as creation and deletion of directories 
could help analysis applications manage data.
Adding file access protocols to object stores
would broaden their use cases and increase their adoption rate.
%


With the emergence of low-cost cloud and on-premise object
storage~\cite{AWS}~\cite{MicrosoftCloud}~\cite{GoogleCloud},
object storage file systems can be a cost effective alternative to block-based 
file systems.
File system interfaces over object storage
would allow for a seamless migration of existing applications~\cite{Inman:2017}.
This would reduce storage costs without requiring
porting applications to a new interface.
%
It would also aid the
current efforts of cloud providers towards storage consolidation to eliminate
{\em storage sprawl}---the spread of data across different media and interfaces.
Storage sprawl causes numerous issues, including 
over-provisioning,
cost inflation, reduced backup efficiency, and 
poor quality of service (QoS)~\cite{DataCenterStorage}.

The key to eliminating storage sprawl is {\em dual access}---the ability to read and write data through both 
file system interfaces and object storage APIs.
However, there are many design considerations that expose 
tradeoffs in the quality and performance of dual access.
We explore these tradeoffs using 
our implementation.

We recommend an initial design for dual access that chooses a simple file-to-object mapping and a
more complex indirect naming scheme.  This design allows data access from the native object API without modifications,
but makes identifying data across systems more complex in order to avoid 
data copies during file system metadata operations.  
Other choices are discussed with quantified tradeoffs.  
We also observe that write-back caching is critical to 
making dual access efficient.
 We conclude that file systems built on object stores can
eliminate storage sprawl while realizing a large fraction of the performance of the underlying object store.

\section{Related work}
\label{section:related}


%
Object storage file systems are not the same as file systems
designed to use object-based storage devices (OSDs), for example
Lustre~\cite{Schwan:2003}.
The object storage file systems that we explore were developed 
to operate over existing, generic object storage systems, and to support data and
application portability.
%
%

%
%
Object storage file systems can be broadly categorized using two properties.
First, whether they are generic and support multiple back-end object stores,
important for providing flexibility and avoiding vendor lock-in.
Second, whether they are compliant with POSIX standards, which we refer to as POSIX complete.
%
%
S3Fuse~\cite{S3Fuse}, Goofys~\cite{Goofys}, RioFS~\cite{RioFS}, GCSfuse~\cite{gcsfuse},
Blobfuse~\cite{AzureFS}, SVFS~\cite{SVFS} and MarFS~\cite{Inman:2017} are some examples of file 
systems which are generic, but not POSIX complete.
Most of these systems do not support POSIX features such as symlinks, hardlinks, or file attributes (chmod),
and have poor performance for random writes.
CephFS~\cite{Weil:2006} is a popular example of a POSIX complete file system;
however, it can only function with its own RADOS object store. 
Some recent systems are both generic and complete, but lack support for dual access.
These do not address the issue of storage sprawl.
S3QL~\cite{S3QL} is POSIX complete and generic but does not support distributed access.
BlueSky~\cite{Vrable:2012} uses caching gateways to aggregate 
writes in log segments that are later pushed to an object store. 
SCFS~\cite{Bessani:2014} is a generic system with near-POSIX semantics that 
supports distributed access.


\section{Design considerations}
\label{section:design}

We believe that dual access to the data through both object and
file-system interfaces is critical to eliminate storage sprawl.
For this paper, we assume a limited, yet representative,
object interface with the following operations:
(1)~\texttt{PUT(name, data)} adds a named object,
(2) \texttt{GET(name)} retrieves an object, and
(3) \texttt{DEL(name)} deletes an object by name.
%
%
%
The file system interface has 
POSIX-defined operations, such as file create, open, read, write, close, delete,
rename, as well as directory operations.
Object and file-system interfaces have a number of fundamental
\emph{namespace} and \emph{data access} differences.
For instance, in object stores
users cannot create directories and subdirectories of objects,
and cannot operate on directories as a whole, 
e.g., rename them.
(Though buckets
of objects are
supported by many object stores they cannot be nested and their number is often limited,
e.g., to around 100 per account in AWS S3~\cite{S3}).
%
%
%
%
Data access differences include the inability of object storage to perform
an in-place partial update of the data, a common operation
in many file system workloads.
%
%

In light of these disparities, providing dual access 
is challenging.
In fact, the goal of dual access is often in direct conflict
with that of achieving high performance.
We present design considerations
for overlaying a file system on generic object storage
and explore the corresponding performance impacts.
In the following text we refer to an implementation of an abstract file system 
working on top of an object store as \emph{ObjectFS}.




\subsection{File-to-object mapping}
\label{sec:mapping}

A fundamental question in the design of ObjectFS is how
to map files to objects.

\ITEM{1$\Rightarrow$1} mapping represents a whole file with a
single object in the object store.
This mapping allows simple and intuitive dual access to the data from the
user perspective.
%
%
1$\Rightarrow$1 mapping can drastically reduce the performance of file writes
because 
%
%
a small modification to a file requires a \texttt{GET}
and a \texttt{PUT} on the complete object.

\ITEM{1$\Rightarrow$N} mapping splits an individual file into multiple objects,
each storing a segment of the file.
The segments can be of a fixed size, as in a traditional block-based file
system, or of variable size, as in an extent-based file system.
Splitting a file into multiple objects enables faster in-file updates
by only writing smaller-sized objects which map to the 
updated parts of the file.
%
%
However, accessing the data from the object interface in 1$\Rightarrow$N
mapping is no longer intuitive and requires additional metadata 
in object-based user applications.

\ITEM{N$\Rightarrow$1} mapping packs multiple files into a single object.
This can improve performance when a subset of small-sized files tend to
be accessed together.
Accessing data through the object interface is even more complicated with
N$\Rightarrow$1 mapping than with 1$\Rightarrow$N mapping.
%


\ITEM{Hybrid} mapping varies the mapping within the
same file system.
For example, ObjectFS could create new objects for each incoming write (as extents) and
then reassemble them into complete objects in the background. 
%
This hybrid mapping trades consistency of object and file system
views of the data for performance.

\subsection{Object naming policy}
\label{sec:naming}

%
%
%

Although the naming of the objects is tightly coupled to file-to-object mapping,
we discuss naming separately to isolate and demonstrate relevant difficulties.
For simplicity, we assume 1$\Rightarrow$1 mapping.

\ITEM{\textsc{file-name}} policy names an object identically to the corresponding file.
Such a policy allows intuitive dual access to the data, but with a
substantial caveat.
Two files with identical names but in different directories cause a conflict in
the flat object namespace.
So, this policy is applicable only in limited scenarios, e.g., when a
read-only file system is deployed on a pre-populated object storage
to perform analytics.

\ITEM{\textsc{file-path}} policy creates an object named after the file's complete path.
This policy is both convenient for dual access and avoids conflicts in the
object storage namespace.
However, a rename of a file requires a \texttt{GET} and a \texttt{PUT}, making
metadata operations slow.
A directory rename requires a
\texttt{GET}-\texttt{PUT} sequence for every file in the directory and performance
scales down as the total size of all files in a directory grow.
%
%
%
%
%
Another limitation of \textsc{file-path} policy is its inability to support
hardlinks (different files referring to the same data).

\ITEM{\textsc{inode-number}} policy names the file using the file system inode number as the object name.
The assumption here is that ObjectFS,
similar to a traditional UNIX file system,
maintains a mapping of file paths to inode numbers.
File paths are translated to inode numbers using a lookup procedure.
The \textsc{inode-number} policy hinders dual access, 
because the inode number needs to be looked up.
Similarly, files created through the object store will need to reference
file system metadata for a name.
%
%
\textsc{inode-number} performs renames quickly as no objects need to
be moved: only the mapping is updated.

%

%
%
%
%

%

\ITEM{\textsc{user-defined}} policies allow the user to drive the naming scheme.
%
%
%
In one potential implementation,
the inode in ObjectFS records the name of the corresponding object.
When a new file is created, ObjectFS executes a user-defined naming policy 
to derive the object name.
The naming policy can take as an input such file system information as the 
file name and path, owner, inode number, and more.
%
%
A corresponding naming policy is required to generate full file paths based 
on the properties of any objects created directly in the object store.
\textsc{user-defined} policies are more flexible than \textsc{inode-number} but
need to be carefully designed to be convenient,
and avoid naming conflicts.

\subsection{Metadata}
\label{sec:metadata}

%
%
%
%

ObjectFS could potentially use several different locations to persistently store its metadata.

\ITEM{\textsc{in-object}} placement stores file metadata (e.g., owner,
permissions, timestamps) in the object store itself.
One option is to store metadata in the same object as data, but this requires
cloud-native applications to deal with metadata during \texttt{GET}s and
\texttt{PUT}s, which compromises dual access.
Another option maintains separate metadata objects: one per file or one
per a group of files.
In this case, dual access is not directly hindered but ``confusing'' metadata
objects are visible in the results of a \texttt{LIST} request.
%
%
%
%
%
Furthermore, object stores typically exhibit high latency, which would
metadata operations (e.g., accessing or updating \emph{atime} or
\emph{uid}).
This typically leads to poor overall performance. 
%
%

\ITEM{\textsc{in-object-meta}} relies on the fact that the majority of object
storage implementations can store user-defined metadata in association with an
object.
Access to user-defined metadata is independent of access to the
object data, and has comparatively lower latency.
The concept is similar to extended attributes in file systems.
%
%
%
This approach offers for dual \emph{metadata} access in addition to dual \emph{data} access.
Object-based applications can request file-system metadata 
through the object interface. However, it relies of a richer object API that is not generic.

\ITEM{\textsc{independent}} stores the file system metadata in a storage
solution separate from the object store.
A key-value store with high scalability and low access latency is one feasible 
configuration.
In this case, metadata operations like inode lookup or
\texttt{stat()} would not require slow accesses to the object storage.
A downside is the higher system complexity and the need to maintain
additional storage system for metadata.
%
%
\textsc{independent} precludes accessing metadata through the object
interface.  Although, ObjectFS could asynchronously write metadata from
the metadata store to corresponding objects in one of the other formats.
%

%
%
%


\subsection{Caching}

%
Caching plays an integral role in file system performance.
For ObjectFS, both read and write caches are important because
the underlying object storage has high latencies and operates efficiently
only when transferring large objects.
%
%
%
%
We limit our discussion to two fundamental design options:

\ITEM{\textsc{local}} cache has its independent instances on every node where the file
system is mounted.
Each cache instance buffers data read or written by the local node.
RAM or a local SSD can be used for cache space.
For \textsc{local} cache, ObjectFS needs to maintain cache consistency between
nodes using, e.g., lock tokens~\cite{schmuck2002gpfs}.
%
%
%
Since object-based accesses do not go through the file system cache,
cloud-native applications could see outdated versions of the data
until caches are synced.

\ITEM{\textsc{unified}} cache is a distributed and shared tier between 
file system clients.
Data cached by one client can quickly be fetched from the cache by other
clients.
%
%
%
%
Redis~\cite{Sanfilippo:2009} and Memcached~\cite{Fitzpatrick:2004} are systems suitable to 
implement a \textsc{unified} cache. 
Caching nodes can be collocated with file system mount nodes or deployed in a 
separate cluster.
A \textsc{unified} cache may re-export an object interface so that object-based
applications access the same data consistently and realize the benefits of the
cache.

\section{Implementation}
\label{section:implementation}

\begin{figure}[t]
  \includegraphics[width=\linewidth]{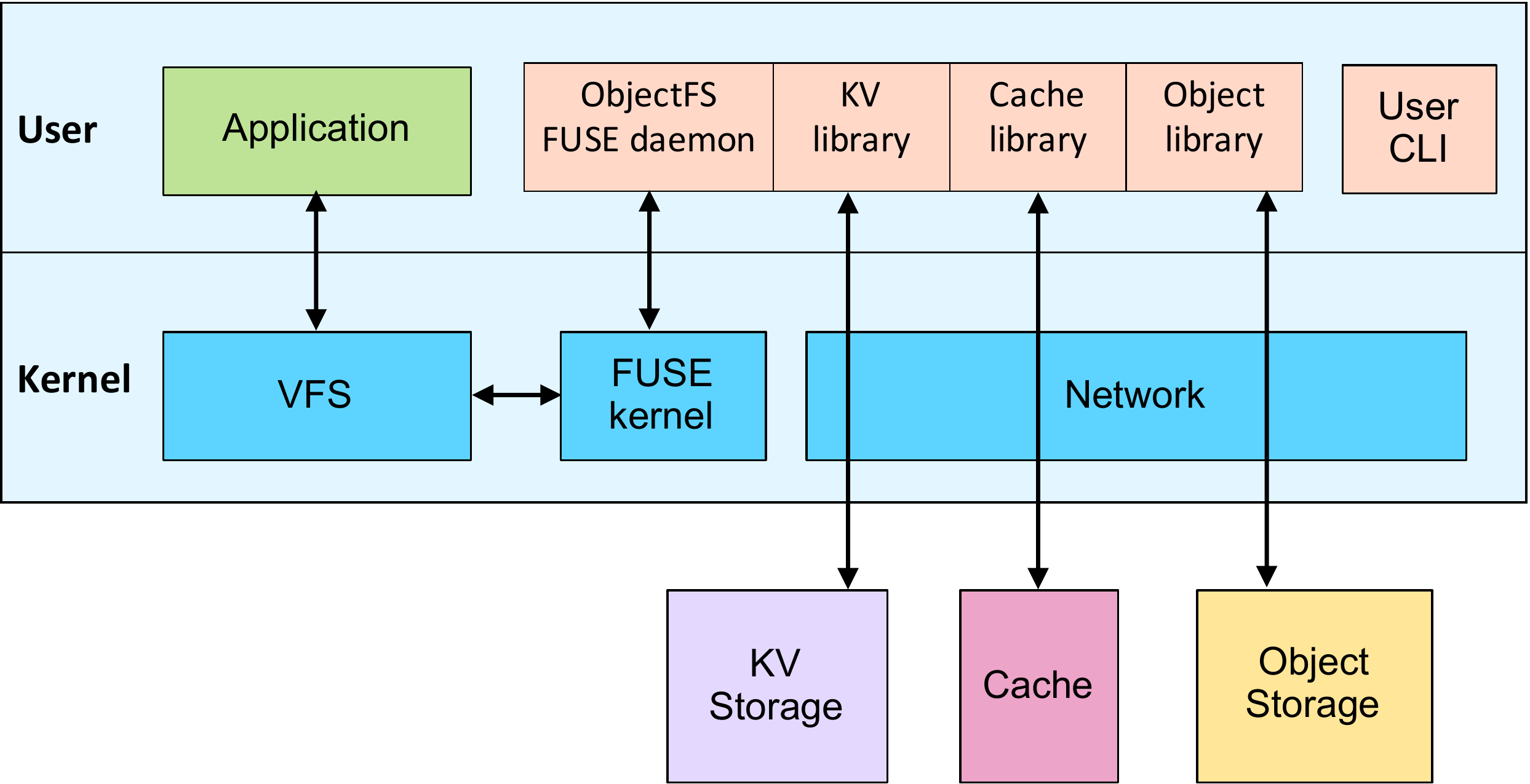}
  \caption{An overview of ObjectFS and the data flow across different components.}
  \label{fig:layers}
\end{figure}

To illustrate the design choices of Section
\ref{section:design}, we developed an ObjectFS prototype using
FUSE~\cite{Vangoor:2017}.
Our implementation is simple and modular to facilitate experimentation with various
ObjectFS configurations.
Figure~\ref{fig:layers} depicts ObjectFS's  high-level architecture.
%
%
%
%
%
%
%
%
ObjectFS's user-space daemon is responsible for the main logic of the file
system: to perform file lookups, reads, writes, etc.
ObjectFS uses an independent metadata service that is
abstracted as a key-value store. 
In this paper we use Redis, an \emph{in-memory} key-value store, because of
its low latency, distributed design, support of transactions, and ability to
persistently store in-memory data.

The object library communicates with object storage using
a common subset of object operations.
We currently support AWS S3~\cite{S3} and OpenStack Swift~\cite{Swift}.
Many object stores support multi-part upload and download of 
large objects.
If available, ObjectFS utilizes this feature to improve performance.
%
%
%
ObjectFS can be configured to run with or without a cache.
Currently we use Redis to cache data.
%
By default, 
data is fetched into the cache on a file open and is flushed back to the object
store on a file close. 
%
%

ObjectFS supports all major POSIX operations, and we are able
to successfully boot a Linux OS directly from an object store
using ObjectFS.
Our implementation is open source and is available for collaborative development
and reuse at our GitHub repository
(\url{https://github.com/objectfs/objectfs}).

%
%
%
%

\section{Evaluation}
\label{section:evaluation}

Our evaluation quantitatively demonstrates some key design
trade-offs presented in Section~\ref{section:design}.
%
%
%
%
%
%
We chose four basic workloads---streaming reads, streaming writes, random
writes, and renames---and measured performance on various ObjectFS
designs.
We used Amazon Web Services (AWS) as a testbed~\cite{AWS}.
An ObjectFS client was mounted on a \textit{t2.2xlarge} compute instance with 8
vCPUs, 32GB of RAM, and \textit{AWS moderate} network bandwidth.
%
ObjectFS's metadata server was deployed on the same instance.
We used AWS S3 object storage with the default standard class of storage
as a backend.
The S3 buckets had default settings with object logging enabled  and
object versioning and transfer acceleration disabled.
%

\begin{figure}[ht]
  \includegraphics[width=\linewidth, height=0.8\linewidth]{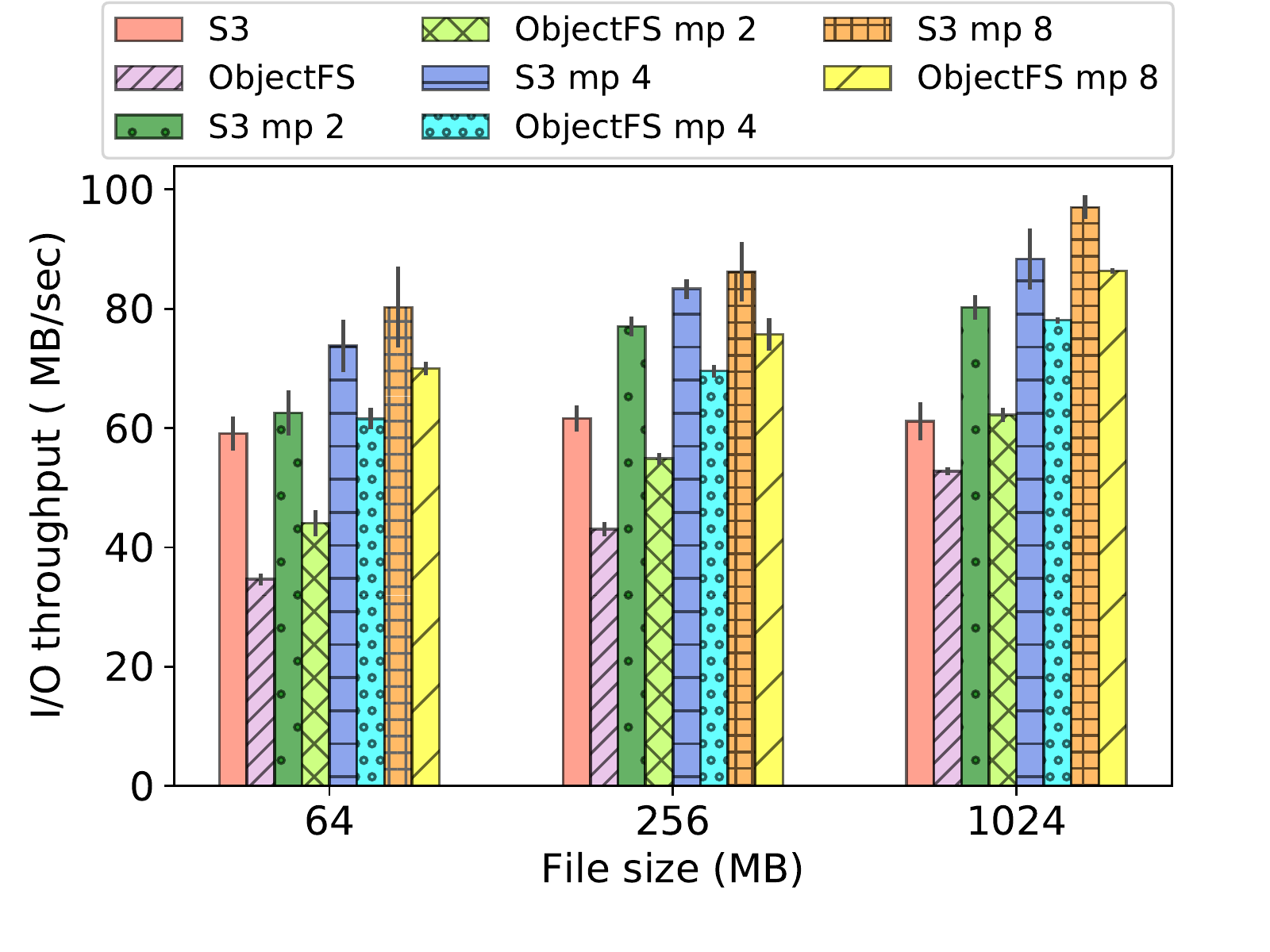}

  \vspace{-5pt} 

  \caption{Streaming read performance on native S3 and ObjectFS. \emph{mp} represents
  multipart download with the corresponding number of threads used.}
  \label{fig:streaming_read}
\end{figure}

\noindent{\bf Streaming reads:}
Read experiments demonstrate that ObjectFS tracks the performance
of the underlying object store for sequential workloads.
We perform sequential reads, in 4 MB record sizes,
on files stored as objects in 1$\Rightarrow$1 mapping.
%
%
We measure the I/O throughput of S3 and ObjectFS when 
varying the file size from 64MB to 1GB and using S3's multipart download with 
2, 4, and 8 threads.
%
%
Multipart downloads divide the object access into parts, parallelized
over multiple threads.
%
%
%
%
%
Figure \ref{fig:streaming_read} shows that multipart downloads
mitigate the performance overhead of ObjectFS and that ObjectFS
realizes a large fraction of S3's potential bandwidth.  
The small remaining overhead comes from metadata operations and caching overhead.
%
%
%
%
%
%

%


\begin{figure}[ht]
  \includegraphics[width=\linewidth, height=0.8\linewidth]{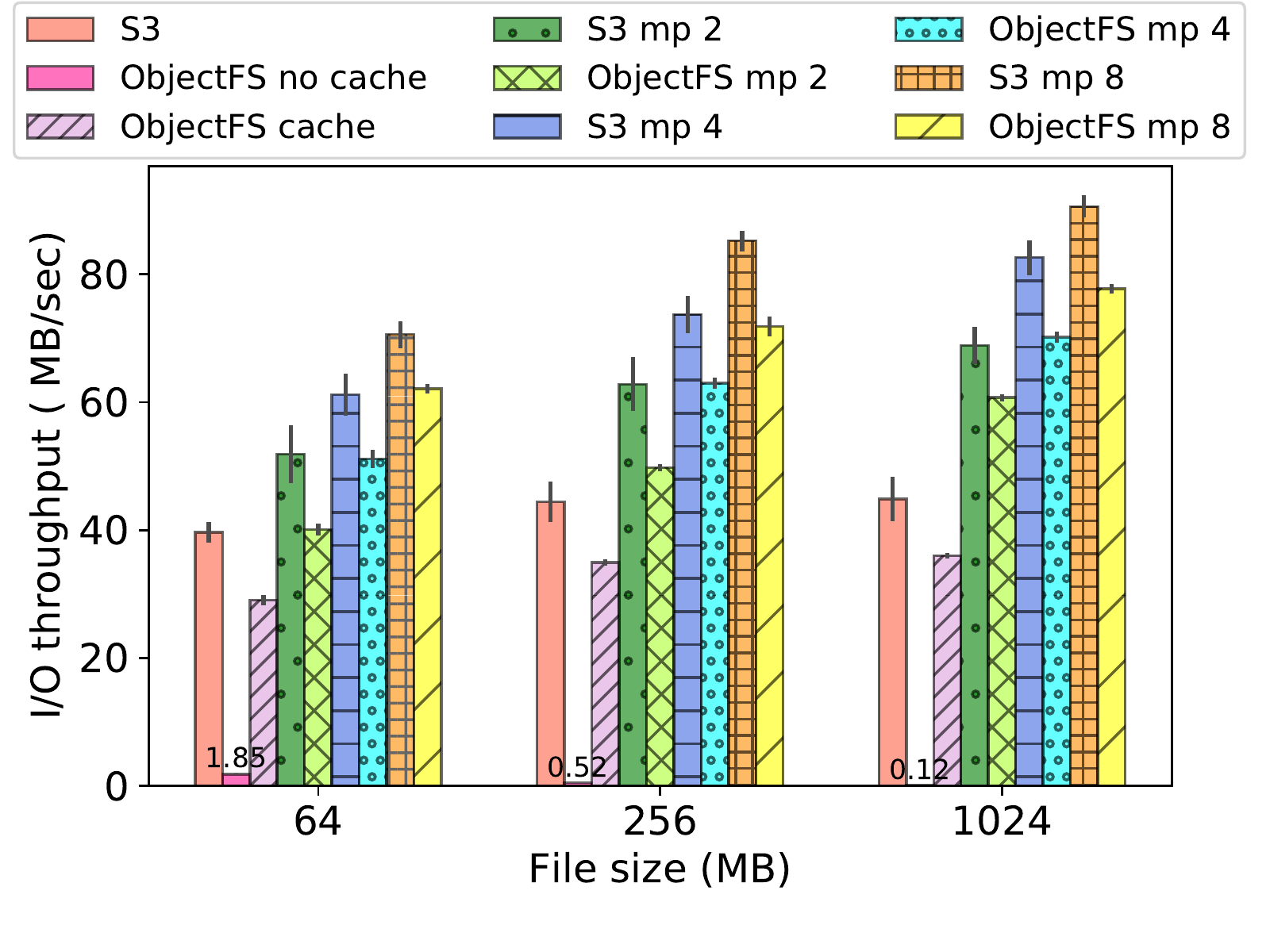}

  \vspace{-5pt}

  \caption{Streaming write performance on native S3 and ObjectFS. \emph{mp} represents multipart
  upload with the corresponding number of threads used.}
  \label{fig:streaming_write}
\end{figure}

\noindent{\bf Streaming writes:}
%
%
Write experiments demonstrate that
caching is critical to realizing performance in ObjectFS.
The experiment performs sequential writes, in 4 MB record sizes,
on files stored as objects in 1$\Rightarrow$1 mapping.
We measure the I/O throughput of S3 and ObjectFS when varying the file size 
from 64MB to 1GB and using S3's multipart upload with 2, 4, and 8 threads.
%
%
%
ObjectFS implements a write-back cache: a write is stored locally in 
a Redis memory store and written back to the object store when the object is closed.
%
%
%
%

Figure \ref{fig:streaming_write} shows that caching enables reasonable 
write throughput when compared with native S3 bandwidth, and that multi-part 
uploads reduce overhead and increase throughput.  Multi-part uploads 
overlap data transfer with metadata operations in multiple threads.
Without a write-back cache, each write results in a read-modify-write in the object store,
reducing throughput to less than 2MB/sec. 
With caching, we aggregate writes in the cache and issue many transfers in parallel.
%
%
%
Increasing multi-part uploads beyond 8 threads shows no more performance improvement.
We theorize that the 
physical footprint of an object is limited to a few storage servers and that more threads result 
in smaller messages to the same set of servers.
%


\begin{figure}
  \includegraphics[width=\linewidth, height=0.7\linewidth]{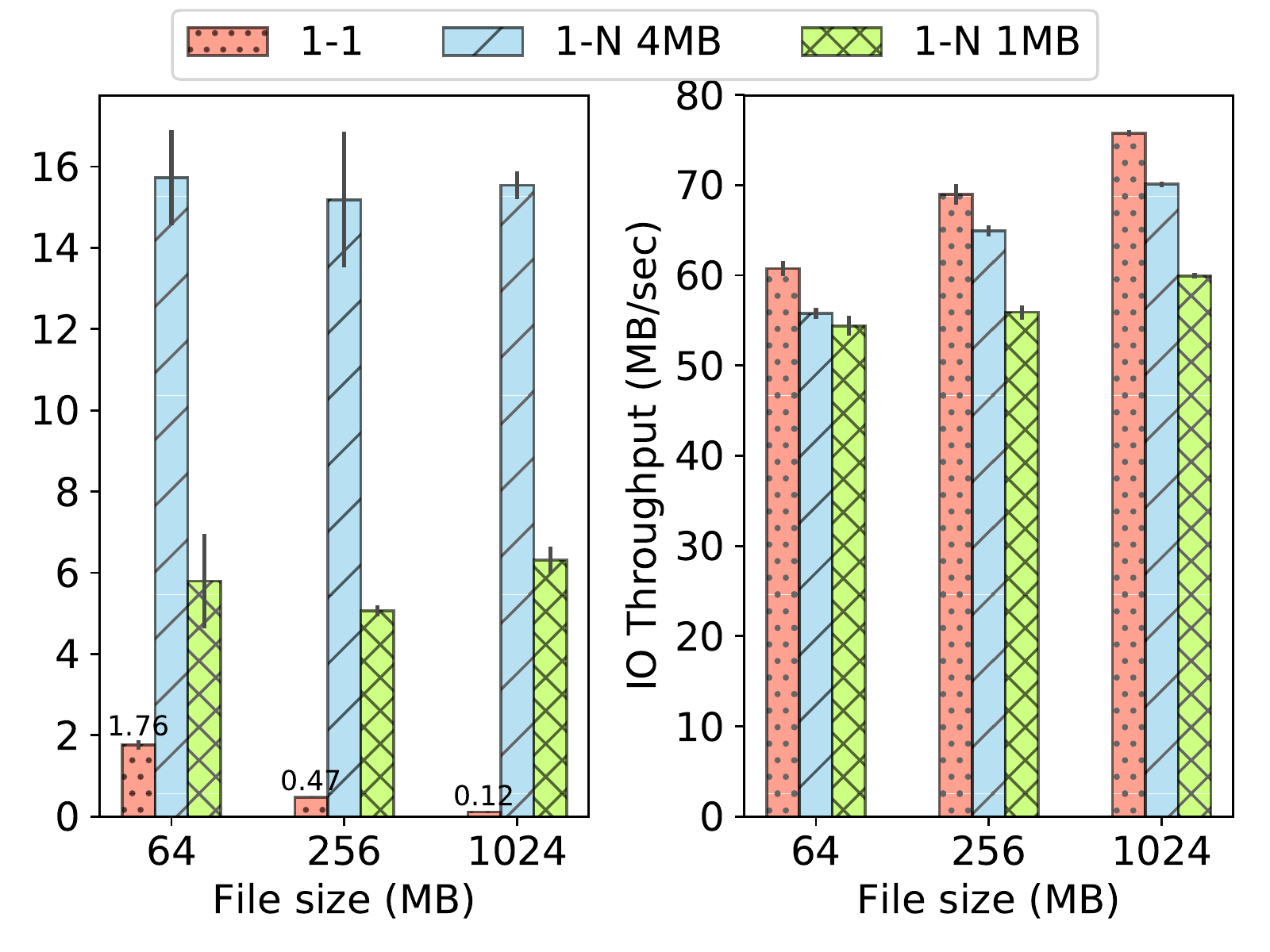}
  \caption{Random write performance for different file object mapping designs without no caching (left) and with 
write-back caching (right).}
  \label{fig:random}
\end{figure}


\noindent{\bf Random writes:}
%
%
Evaluating the performance of random writes demonstrates the performance tradeoffs 
among the different file to object mappings.
%
We perform writes of 4 MB to random file offsets aligned to 4 MB using different mapping schemes: 1$\Rightarrow$1, 1$\Rightarrow$N (1 MB chunks)
and 1$\Rightarrow$N (4 MB chunks).  
Figure~\ref{fig:random} (left) shows throughput for write-through workloads without caching.
In this scenario, 1$\Rightarrow$1 is much worse than 1$\Rightarrow$N, 1.7 MB/s versus 15 MB/s.  
With 1$\Rightarrow$1 mapping, each 4 MB write performs a partial write or read-modify-write against the underlying
object, whereas 1$\Rightarrow$N mappings write entire object(s).
More importantly, all data rates are remarkably low without caching. 

With write caching, Figure~\ref{fig:random} (right), ObjectFS defers individual writes and avoids read-modify-writes to
realize an order-of-magnitude performance improvement.
This experiment uses 8 I/O threads, doing multi-part upload for 1$\Rightarrow$1 and parallel transfers to mulitple objects for 1$\Rightarrow$N.
The 1$\Rightarrow$N mappings are slightly slower than 1$\Rightarrow$1 due to overhead for RESTful calls to more objects.
Caching raises the random-write throughput of ObjectFS close
to the sequential performance of S3.

%





\begin{figure}
  \includegraphics[width=\linewidth, height=0.7\linewidth]{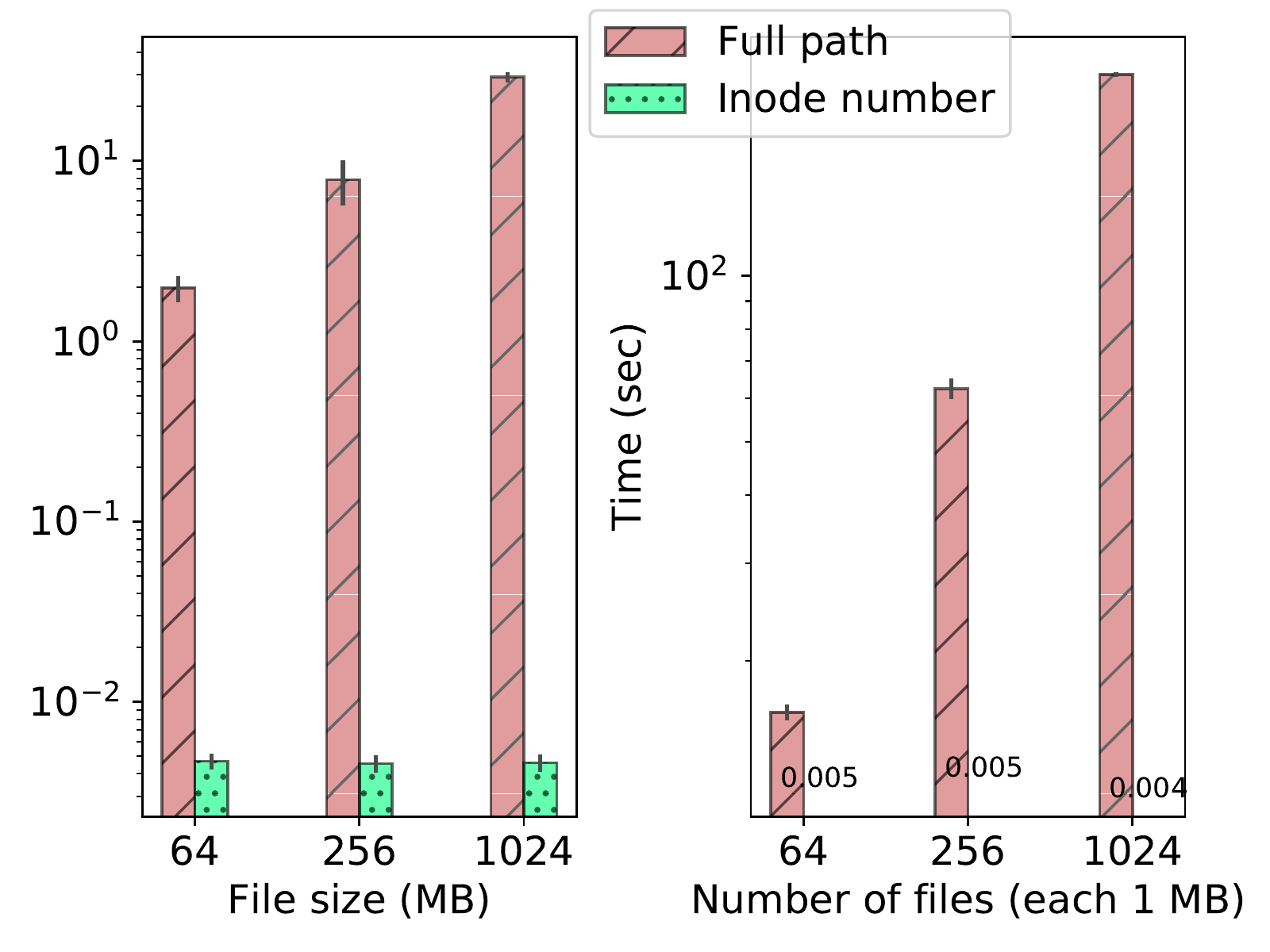}
  \caption{Rename latency for different object naming designs.  Experiments look at renaming a single large file (left) and many 1 MB files (right).}
  \label{fig:rename}
\end{figure}

\noindent{\bf Metadata performance:}
We also examine the performance associated with different file naming conventions that
affect dual access.   We perform two experiments:  The first, 
Figure~\ref{fig:rename} (left), renames files of different sizes.
With full path naming, a rename results in an S3 server-side copy of the object; 
the performance of rename operations thus scales with the file size, taking 
2 seconds for a 64MB file and 30 seconds for 1GB file.
When naming by inode number,
rename is fast (less than 0.005 seconds) and does not depend on file size;
in this case renames in ObjectFS are metadata-only operations.
The second experiment, Figure~\ref{fig:rename} (right), renames directories with varying number of files, each 1MB in size.
With full path naming, latency increases with the number of files.  
With inode naming, latency is consistent and low.

\section{Discussion and Conclusions}

Dual access to data through object and file system APIs is feasible with
a judicious choice of design options.
Based on our evaluation, we propose a specific design that 
preserves object APIs and incurs only minor overheads when accessing data through 
the ObjectFS file system.  Specifically, a 1$\Rightarrow$1 file to object mapping
allows the object store APIs to access data without assembling data from multiple objects.
We recommend an indirect naming scheme based on naming objects by file system inode number.
This choice enables the system to perform metadata operations without copying, but adds
complexity to object access, which must resolve the file system name to an inode number.
This is only one design; our evaluation quantifies tradeoffs and thus can
aid in future designs of object storage file systems.

We also conclude that write-back caching is a critical
technology for deploying object-based file systems.
Caching aggregates multiple writes in memory, converting many synchronous writes
into fewer larger asynchronous writes.  Without caching, object file systems have
low throughput and high latency. This would limit the applications that could adopt
object based file systems to those that perform synchronous writes infrequently.

%
%

%
%
%
%
%
%

\vspace{5pt}
\noindent\textbf{Acknowledgements:} This work was funded by a study grant from IBM Research.

\balance

{\footnotesize \bibliographystyle{acm}
\bibliography{objectfs2018}

\IfFileExists{./objectfs2018.ent}{
  \theendnotes
}{
}

\end{document}